\documentclass[aps,prd,showpacs,nofootinbib,floatfix,eqsecnum]{revtex4}  
\usepackage{amsmath,latexsym,graphicx,multirow,longtable}
\usepackage{verbatim,enumerate}

\newcommand{\pder}[2]{\frac{\partial #1}{\partial #2}}

\begin{document}
\title{Multipole moments of bumpy black holes}
\author{Sarah J.\ Vigeland}
  
\affiliation{Department of Physics and MIT Kavli Institute, MIT, 77 Massachusetts Ave., Cambridge, MA 02139}

\date{\today}
\begin{abstract}
General relativity predicts the existence of black holes, compact objects whose spacetimes depend on only their mass, spin, and charge in vacuum (the ``no hair'' theorem). As various observations probe deeper into the strong fields of black hole candidates, it is becoming possible to test this prediction. Previous work suggested that such tests can be performed by measuring whether the multipolar structure of black hole candidates has the form that general relativity demands, and introduced a family of ``bumpy black hole'' spacetimes to be used for making these measurements. These spacetimes have generalized multipoles, where the deviation from the Kerr metric depends on the spacetime's ``bumpiness.'' In this paper, we show how to compute the Geroch-Hansen moments of a bumpy black hole, demonstrating that there is a clean mapping between the deviations used in the bumpy black hole formalism and the Geroch-Hansen moments. We also extend our previous results to define bumpy black holes whose {\it current} moments, analogous to magnetic moments of electrodynamics, deviate from the canonical Kerr value.
\end{abstract}

\maketitle

\section{Introduction}
\label{sec:intro}
In many areas of physics, multipolar expansions are used as tools for describing the shape of a distribution of matter or energy, or for describing the behavior of a potential function. Multipole moments are most commonly used to describe fields whose governing equations are linear, since the functions describing the angular behavior are typically eigenfunctions of the angular piece of the governing differential operator. For example, the Newtonian gravitational potential $\Phi$ arising from a matter distribution $\rho$ must satisfy Poisson's equation:
\begin{eqnarray}
	\nabla^2\Phi = \left\{ \begin{array}{lr} 4\pi G \rho & \mathrm{(interior),} \\ 0 & \mathrm{(exterior).} \end{array} \right.
\end{eqnarray}
In the exterior region, we can write $\Phi$ as a sum over multipolar contributions:
\begin{eqnarray}
	\Phi(r,\theta,\phi) = - G \sum_{lm} \frac{M_{lm}Y_{lm}(\theta,\phi)}{r^{l+1}} \;.
	\label{Newtonian_multipole}
\end{eqnarray}
The coefficients $M_{lm}$ are mass multipole moments. By matching the expansion of $\Phi$ on the boundary to a similar expansion for the interior, they can be shown to describe the angular distribution of the mass of the source. (Throughout this paper, we will restrict ourselves to axisymmetric spacetimes, for which the axial index $m$ must be zero; we ignore it in what follows.)

Using multipole moments to describe gravity in general relativity is not as simple, thanks largely to the nonlinear nature of the governing equations. Geroch {\cite{Geroch}} and Hansen {\cite{Hansen}} developed a very useful multipolar description for spacetimes of isolated, stationary, axisymmetric objects in GR in terms of scalar multipoles. Their definition applies to spacetimes that are asymptotically flat; for such spacetimes there is a well-defined ``large $r$'' limit in which multipoles can be defined in a way that roughly accords with our usual intuition. When one computes the Geroch-Hansen moments of a source, one finds that its spacetime is described by a family of mass moments $M_l$, very similar to those appearing in our Newtonian expansions in Eq.\ (\ref{Newtonian_multipole}), as well as a family of {\it current} moments $S_l$. For a fluid body, the current moments describe how the matter flow is distributed through the compact body, much as magnetic moments describe how electric current is distributed through an electromagnetic source. These moments can be conveniently combined in the complex moment ${\cal M}_l = M_l + i S_l$.

For a generic source, the moments ${\cal M}_l$ are unconstrained.\footnote{If we assume the spacetime is reflection symmetric, then only the even mass moments and odd spin moments can be nonzero. In this paper, we do not restrict ourselves to reflection symmetric spacetimes.} As we will discuss in more detail in the following section, for a Kerr black hole the moments take a particularly simple form:
\begin{eqnarray}
	\mathcal{M}_l = M(ia)^l \;,
	\label{nohair}
\end{eqnarray}
where $\mathcal{M}_l$ is the $l$th moment, $M$ is the total mass of the black hole, $a$ is its spin parameter, and we are using units where $G=c=1$. (We neglect the astrophysically uninteresting possibility of a black hole with macroscopic charge.) This is a statement of the ``no-hair'' theorem: the spacetime of a Kerr black hole is completely described by its mass and spin \cite{nohair}.

The sharply constrained nature of the Kerr multipoles implied by Eq.\ (\ref{nohair}) suggests that this relation may be useful as a test of black hole spacetimes: if the spacetime is Kerr, then knowledge of only two moments is needed to determine all of the others. Ryan {\cite{Ryan}} was the first to build a scheme to test this idea, constructing spacetimes in which all of the moments were arbitrary. This scheme was sufficient to prove the principle of the idea, but did not work well for building spacetimes good deep into the strong field. Collins and Hughes {\cite{CH}} noted that, if general relativity correctly describes black hole candidates, then testing their nature amounts to trying to falsify the hypothesis that they are Kerr black holes. They suggested formulating black hole tests as a null experiment by examining spacetimes for which
\begin{eqnarray}
	\mathcal{M}_l = M(ia)^l + \delta\mathcal{M}_l \;,
\end{eqnarray}
and using measurements to test whether $\delta{\cal M}_l=0$, as it should if they are Kerr black holes.

Collins and Hughes formulated a simple version of this test by showing how to deform the multipoles of a non-rotating black hole. Glampedakis and Babak {\cite{GB}} extended this idea by modifying the quadrupole moment of a Kerr black hole. Though not as generic in the form of the moment that can be modified, this significantly improved the astrophysical relevance of this test by allowing for spacetimes with angular momentum. Vigeland and Hughes {\cite{VH}} then improved the Collins and Hughes construction, using the Newman-Janis algorithm {\cite{NJ}} to transform bumpy Schwarzschild black holes into bumpy Kerr black holes. Their approach allows one to vary any of the spacetime's mass moments, an important consideration since black hole candidates may agree with the Kerr metric's multipoles up to some $l_{\rm max}$ but differ for $l > l_{\rm max}$.

We have chosen to focus on the Geroch-Hansen moments, as opposed to the tensor moments described by Thorne {\cite{Thorne1980}}, for the sake of simplicity. However, as we discuss in Sec.\ \ref{sec:GH_moments}, Geroch-Hansen moments are only defined for spacetimes for which $R_{\mu\nu}=0$. As we show in Sec.\ \ref{sec:KerrBBH_mass} and \ref{sec:KerrBBH_spin}, when we add bumps to the Kerr metric, the spacetime is no longer vacuum in GR. It will turn out that for the perturbations we have considered, the nonvacuum nature is of sufficiently ``high order'' (in a sense made precise in Sec.\ \ref{sec:KerrBBH_mass}) that we can still define Geroch-Hansen moments up to some order. In order to study the multipolar structure of other kinds of perturbations to the Kerr metric, we would need to calculate the spacetime's tensor moments. For example, this would facilitate analyzing the ``bumpiness'' of black holes in other theories of gravity, such as dynamical Chern-Simons extensions to general relativity. We leave this for future work.

The purpose of this paper is twofold. First, we map the mass perturbations described in {\cite{VH}} to perturbations in the Geroch-Hansen moments $\delta\mathcal{M}_l$. In constructing a bumpy black hole, we work in a few different coordinate systems, and specify the spacetime's bumps by adding a function which is proportional to a spherical harmonic in one of those coordinate systems. Our goal is to demonstrate that this choice maps in a natural way to the spacetime's Geroch-Hansen moments. For example, a bump which is built from an $l=2$ spherical harmonic changes the even Geroch-Hansen moments above the $l=2$ moment. In principle, one could construct a change to a single Geroch-Hansen mass moment by including multiple appropriately weighted spherical harmonic terms in the bumpy black hole spacetime.

Second, we show how one can build bumpy black hole spacetimes for which the current moments are modified. This expands the domain of tests that one can perform, making it possible to test the full range of the gravitational moments. For example, Yunes and Pretorius {\cite{YP}} showed that dynamical Chern-Simons extensions to general relativity lead to a rotating black hole solution that is not the Kerr metric; in the slow rotation limit, the spacetime's moments are the same as Kerr for $l < 3$, but its $l=4$ current moment differs from the Kerr value {\cite{SY}}. Having control over both a spacetime's mass and spin moments completes our ability to build bumpy black holes and allows us to test the Kerr metric in great detail. We show that the spin moment perturbations behave similarly to the mass moment perturbations; if we add an order $l$ spin moment perturbation, which we define in Sec.\ \ref{sec:SchwBBH_spin} and \ref{sec:KerrBBH_spin}, we leave the spin moments unchanged up to order $l$. When we allow for both mass moment and spin moment perturbations, the bumpy Kerr spacetime is described by three perturbation potentials $\psi_1$, $\gamma_1$, and $\sigma_1$:
\begin{eqnarray}
	\nonumber ds^2 &=& -\left[(1+2\psi_1) \left(1-\frac{2Mr}{\Sigma}\right) + \frac{4aMr}{\Sigma^2} \sigma_1(r,\theta) \right] dt^2 - \gamma_1 \frac{4a^2Mr\sin^2\theta}{\Delta\Sigma}\:dt\:dr \\
	\nonumber && - \left[ (1+2\psi_1-\gamma_1)\frac{4aMr\sin^2\theta}{\Sigma} + 4\sigma_1 \left(\frac{(r^2+a^2)2Mr}{\Sigma^2}+\frac{\Delta}{\Sigma}-\frac{\Delta}{2\Sigma-4Mr}\right) \right]\:dt\:d\phi \\
	\nonumber && + (1+2\gamma_1-2\psi_1) \frac{\Sigma}{\Delta} dr^2 + 2\gamma_1 \left[ 1 + \frac{2Mr (r^2+a^2)}{\Delta\Sigma} \right] a \sin^2\theta\:dr\:d\phi + (1+2\gamma_1-2\psi_1)\Sigma\:d\theta^2 \\
	\nonumber && + \left[ (r^2+a^2)^2 - a^2 \Delta\sin^2\theta + (\gamma_1-\psi_1) \frac{8a^2M^2r^2\sin^2\theta}{\Sigma-2Mr} - 2\psi_1\frac{\Sigma^2\Delta}{\Sigma-2Mr} \right.\\
	&& \left. + \frac{4aMr}{\Sigma-2Mr}\sigma_1 \left(\Delta + \frac{2Mr(r^2+a^2)}{\Sigma}\right) \right] \frac{\sin^2\theta}{\Sigma} \:d\phi^2 \;,
\end{eqnarray}
where $\Sigma = r^2+a^2\cos^2\theta$, $\Delta = r^2-2Mr+a^2$, and we are using Boyer-Lindquist coordinates. In the following sections, we will treat mass perturbations and spin perturbations separately; we include the complete bumpy Kerr metric here for completeness.

The remainder of this paper is organized as follows. In Sec.\ \ref{sec:GH_moments}, we review in detail how Geroch-Hansen moments are calculated, and demonstrate the procedure on the Kerr spacetime. We then apply this procedure to the Schwarzschild spacetime with mass perturbations in Sec.\ \ref{sec:SchwBBH_mass}, followed by a Kerr spacetime with mass perturbations in Sec.\ \ref{sec:KerrBBH_mass}. Finally, we present our procedure for perturbing the spin moments of both Schwarzschild (Sec.\ \ref{sec:SchwBBH_spin}) and Kerr (Sec.\ \ref{sec:KerrBBH_spin}) black holes. We conclude in Sec.\ \ref{sec:conclusion} by discussing some directions for future work in this problem. We adopt the following conventions in this paper. We work in geometrized units where $G=c=1$.  When writing tensors, we use a Greek letter to indicate a spacetime index, and a Latin letter to indicate a spatial index.

\section{Computing Geroch-Hansen moments}
\label{sec:GH_moments}
We begin our analysis by showing how to compute the Geroch-Hansen moments for the spacetime of a compact object. As an important example, we demonstrate the procedure on the Kerr metric, which in Boyer-Linquist coordinates is written
\begin{eqnarray}
	ds^2 &=& -\left(1-\frac{2Mr}{\Sigma}\right)dt^2 - \frac{4aMr\sin^2\theta}{\Sigma}\:dt\:d\phi + \frac{\Sigma}{\Delta} dr^2 + \Sigma\:d\theta^2 + \left[ (r^2+a^2)^2 - a^2 \Delta\sin^2\theta \right]\:\frac{\sin^2\theta}{\Sigma}\:d\phi^2 \;.
	\label{Kerr}
\end{eqnarray}

In order to compute the Geroch-Hansen moments, the spacetime must have a timelike Killing vector and be asymptotically flat. Let the spacetime's Killing tensor be $K^\alpha$, and let the manifold $V$ be the 3-surface orthogonal to this vector. The metric on $V$ can be written
\begin{eqnarray}
	h_{ij} = \lambda g_{ij} + K_i K_j \;,
\end{eqnarray}
where $\lambda = -K^\alpha K_\alpha$ is the norm of $K^\alpha$. To calculate the moments, we perform a conformal transformation to map infinity onto a point $\Lambda$. The space is  asymptotically flat if it can be conformally mapped to a 3-space $\tilde{V}$ which satisfies \cite{Geroch}
\begin{enumerate}[(i)]
	\item $\tilde{V} = V \cup \Lambda$, where $\Lambda$ is a single point;
	\item $\tilde{h}_{ij} = \Omega^2 h_{ij}$ is the conformal metric;
	\item $\left.\Omega\right|_\Lambda = 0$, $\left.\tilde{D}_i\Omega\right|_\Lambda = 0$, $\left.\tilde{D}_i\tilde{D}_j\Omega\right|_\Lambda = 2 \tilde{h}_{ij}$;
\end{enumerate}
where $\Omega$ is the conformal factor and $\tilde{D}_i$ is the derivative operator associated with $\tilde{h}_{ij}$. The conformal metric has the form {\cite{BH}}
\begin{eqnarray}
	ds^2 = dr^2 + r^2 d\theta^2 + r^2\sin^2\theta \ e^{-2\beta(r,\theta)} d\phi^2 \;. \label{conformal_metric}
\end{eqnarray}
The function $\beta(r,\theta)$ parametrizes the deviation of the conformal metric from sphericity.

In order to construct the spacetime's Geroch-Hansen moments, we need its Ernst potential {\cite{Ernst}} and the conformal factor $\Omega$. We begin with the Ernst potential, which we build from the norm $\lambda$ and twist $\omega$ of $K^\alpha$. For the Kerr metric, the norm is given by
\begin{eqnarray}
	\lambda &=& 1-\frac{2Mr}{\Sigma} \;.
\end{eqnarray}
The twist is related to the ``generalized curl''\footnote{Normally, the curl of a vector is only defined in a 3-dimensional vector space; we follow the lead of Ref.\ {\cite{FHP}} in generalizing the notion of the curl here.} of the timelike Killing vector,
\begin{eqnarray}
	\omega_\alpha = \epsilon_{\alpha\beta\gamma\delta} K^\beta \nabla^\gamma K^\delta \;.
\end{eqnarray}
From the Bianchi identities, we can write the curl as {\cite{Geroch1971}}
\begin{eqnarray}
	\nabla_{[\alpha} \ \omega_{\beta]} = -\epsilon_{\alpha\beta\gamma\delta} K^\gamma {R^\delta}_\nu K^\nu \;.
\end{eqnarray}
For a spacetime that is vacuum in GR, the condition $R_{\mu\nu}=0$ implies that $\nabla_{[a} \ \omega_{b]}=0$. This allows us to write $\omega_\alpha = \nabla_\alpha \omega$, where the scalar function $\omega$ is the twist of $K^\alpha$. (If $R_{\mu\nu} \neq 0$, we cannot construct the Ernst potential. We will discuss this issue in more detail in Sec.\ \ref{sec:KerrBBH_mass} and \ref{sec:KerrBBH_spin}.) For the Kerr spacetime, we find
\begin{eqnarray}
	\omega &=& -\frac{2Ma\cos\theta}{\Sigma} \;.
\end{eqnarray}
We then combine the norm and twist into the complex quantity
\begin{eqnarray}
	\epsilon &=& \lambda + i \omega \\
		&=& 1-\frac{2Mr}{\Sigma} - i \frac{2Ma\cos\theta}{\Sigma} \;,
\end{eqnarray}
where on the second line we have specialized to Kerr. From this, two definitions of the Ernst potential appear in the literature: Ref.\ {\cite{Ernst}} defines it as
\begin{eqnarray}
	\Xi &=& \frac{1+\epsilon}{1-\epsilon} \;, \label{Ernst_Xi}
\end{eqnarray}
while \cite{FHP} defines the Ernst potential as
\begin{eqnarray}
	\xi &=& \frac{1-\epsilon}{1+\epsilon} \;. \label{Ernst_xi}
\end{eqnarray}
These definitions are simply related to one another ($\xi = \Xi^{-1}$). We find the potential $\xi$ to be most useful for computing multipoles of bumpy black hole spacetimes, but we use $\Xi$ to perturb a spacetime's current multipoles.

Next, we must find the conformal factor $\Omega$. We begin by defining a new radial coordinate $\bar{R}$ according to
\begin{eqnarray}
	r=\bar{R}^{-1} \left(1+M\bar{R}+\frac{M^2-a^2}{4}\bar{R}^2\right) \;. \label{Kerr_Rbar}
\end{eqnarray}
The point $\Lambda$ corresponds to $\bar{R}=0$. The conformal factor is given by
\begin{eqnarray}
	\Omega = \frac{\bar{R}^2}{\sqrt{\left(1-\frac{M^2-a^2}{4}\bar{R}^2\right)^2 - a^2\bar{R}^2\sin^2\theta}} \;,
\end{eqnarray}
and the conformal metric is given by
\begin{eqnarray}
	ds^2 &=& d\bar{R}^2 + \bar{R}^2 d\theta^2 + \bar{R}^2\sin^2\theta \ d\phi^2 \left[1-\left(\frac{4a\bar{R} \sin\theta}{4-(M^2-a^2)\bar{R}^2}\right)^2\right]^{-1} \;.
\end{eqnarray}
This corresponds to $\beta$ equal to
\begin{eqnarray}
	\beta = \frac{1}{2} \ln\left[ 1-\left(\frac{4a\bar{R}\sin\theta }{4-(M^2-a^2)\bar{R}^2}\right)^2 \right] \;.
\end{eqnarray}

As shown by B\"ackdahl and Herberthson {\cite{BH}}, the multipole moments can be computed from derivatives of a function $y$, which we now describe. Begin by defining $\tilde\phi$, a conformally weighted variant of the Ernst potential $\xi$,
\begin{eqnarray}
	\tilde{\phi} = \Omega^{-1/2} \xi \;,
\end{eqnarray}
and new cylindrical coordinates $\tilde z$ and $\tilde\rho$:
\begin{eqnarray}
	\tilde{z} &=& \bar{R} \cos\theta \;, \\
	\tilde{\rho} &=& \bar{R} \sin\theta \;.
\end{eqnarray}
We now write the potential $\tilde\phi$ as a function of these variables, $\tilde\phi(\tilde z,\tilde\rho)$, and introduce yet another variation:
\begin{eqnarray}
	\phi_L(\bar R) = \tilde{\phi}(\bar R, i \bar R) \;.
\end{eqnarray}
We need to define a few more functions related to the metric:
\begin{eqnarray}
	\beta_L(\bar R) &=& \beta(\bar R, i\bar R) \;, \\
	\kappa_L(\bar R) &=& -\ln\left[1-\bar R\int_0^{\bar R} \frac{e^{2\beta_L(\bar R')}-1}{\bar R'^2}d\bar R'-\bar RC\right] + \beta_L(\bar R) \;,
\end{eqnarray}
where $\beta$ is defined in Eq.\ (\ref{conformal_metric}) and $C$ is the integration constant. We can choose the gauge so that $C=0$. The multipoles are calculated from the function
\begin{eqnarray}
	y(\bar R) = e^{-\kappa_L(\bar R)/2}\phi_L(\bar R) \;,
\end{eqnarray}
with the $l$th multipole moment given by
\begin{eqnarray}
	\mathcal{M}_l = \frac{2^l l!}{(2l)!} \left.\frac{d^l y}{d\rho^l}\right|_{\rho=0} \;,
\end{eqnarray}
where $\rho(\bar R) = \bar R e^{\kappa_L(\bar R)-\beta_L(\bar R)}$.

For the Kerr spacetime, the potential $\phi_L$ is given by
\begin{eqnarray}
	\phi_L(\bar R) = \frac{M(1+ia\bar R)}{\left(1+a^2\bar R^2\right)^{3/4}} \;.
\end{eqnarray}
The functions $\beta_L(\bar R)$ and $\kappa_L(\bar R)$ are given by
\begin{eqnarray}
	\beta_L(\bar R) &=& \frac{1}{2}\ln\left[1+a^2\bar R^2\right] \;, \\
	\kappa_L(\bar R) &=& \frac{1}{2}\ln\left[ \frac{1+a^2\bar R^2}{(1-a^2\bar R^2)^2} \right]\;.
\end{eqnarray}
Then the variable $\rho$ is given by $\rho = \bar R (1-a^2 \bar R^2)^{-1}$ and
\begin{eqnarray}
	y(\bar R) = \frac{M\sqrt{1-a^2\bar R^2}}{1-ia\bar R} \;.
\end{eqnarray}
In this case, the multipoles can be written compactly as
\begin{eqnarray}
	\mathcal{M}_l = M(ia)^l \;.
\end{eqnarray}

\section{Perturbations to the mass moments: Schwarzschild background}
\label{sec:SchwBBH_mass}
In this section and the following one, we apply this procedure to compute the multipoles of bumpy black hole spacetimes as presented in Ref.\ {\cite{VH}}. We begin with a bumpy Schwarzschild black hole:
\begin{eqnarray}
	\nonumber ds^2 &=& -(1+2\psi_1) \left(1-\frac{2M}{r}\right)dt^2 + (1+2\gamma_1-2\psi_1) \left(1-\frac{2M}{r}\right)^{-1} dr^2 \\
		&& + (1+2\gamma_1-2\psi_1) \: r^2 \: d\theta^2 + (1-2\psi_1) \: r^2 \sin^2\theta \: d\phi^2 \;.
	\label{SchwBBH}
\end{eqnarray}
The perturbation to the mass moments is described by the potentials $\psi_1$ and $\gamma_1$, with $\psi_1=\gamma_1=0$ corresponding to a Schwarzschild black hole. We further restrict ourselves to perturbations for which the metric satisfies the vacuum Einstein equations to first order; thus the potential $\psi_1$ must satisfy Laplace's equation:
\begin{eqnarray}
	\frac{\partial^2\psi_1}{\partial\rho^2} + \frac{1}{\rho}\frac{\partial\psi_1}{\partial\rho} + \frac{\partial^2\psi_1}{\partial z^2} = 0 \;
	\label{Laplace}
\end{eqnarray}
where the Weyl coordinates ($\rho$, $z$) are related to Schwarzschild coordinates by
\begin{eqnarray}
	\rho &=& r\sin\theta \sqrt{1-\frac{2M}{r}} \;, \label{Weyl_rho} \\
	z &=& (r-M)\cos\theta \;. \label{Weyl_z}
\end{eqnarray}
Because $\psi_1$ satisfies Laplace's equation in Weyl coordinates, we can take it to be a spherical harmonic times an appropriate power of $1/\sqrt{\rho^2+z^2}$. Once we specify $\psi_1$, the potential $\gamma_1$ is calculated by integrating a constraint equation involving the perturbation $\psi_1$ and the background potentials $\psi_0$ and $\gamma_0$, which are given by
\begin{eqnarray}
	\psi_0(r,\theta) &=& \frac{1}{2} \ln \left( 1-\frac{2M}{r} \right) \;, \\
	\gamma_0(r,\theta) &=& - \frac{1}{2} \ln \left( 1+\frac{M^2\sin^2\theta}{r^2-2Mr} \right) \;.
\end{eqnarray}
See \cite{VH} for further details.

The norm and twist of the timelike Killing vector field, to first order in the perturbation, are given by
\begin{eqnarray}
	\lambda &=& (1+2\psi_1) \left(1-\frac{2M}{r}\right) \;, \\
	\omega &=& 0 \;.
\end{eqnarray}
The Ernst potential is given by
\begin{eqnarray}
	\xi &=& \frac{M}{r-M} - \left[1-\left(\frac{M}{r-M}\right)^2\right] \psi_1 \;.
\end{eqnarray}
The conformal factor becomes
\begin{eqnarray}
	\Omega &=& \bar{R}^2 \left(1-\frac{M^2}{4}\bar{R}^2\right)^{-1} \left(1-\gamma_1\right) \;,
\end{eqnarray}
and the conformal metric is
\begin{eqnarray}
	\noindent ds^2 &=& d\bar{R}^2 + \bar{R}^2 d\theta^2 + \left(1-2\gamma_1\right) \bar{R}^2\sin^2\theta \ d\phi^2 \;,
\end{eqnarray}
where $\bar{R}$ has the same definition as for an unperturbed Schwarzschild spacetime,
\begin{eqnarray}
	r &=& \bar{R}^{-1} \left( 1 + M\bar{R} + \frac{M^2\bar{R}^2}{4} \right) \;.
\end{eqnarray}

\subsection{$l=2$ mass perturbation}
Now we calculate the perturbations to the moments for particular choices of $\psi_1$ and $\gamma_1$. We begin by considering a solution to Eq.\ (\ref{Laplace}) that has the form of an $l=2$ spherical harmonic:
\begin{eqnarray}
	\psi_1^{l=2}(\rho,z) &=& \frac{B_2 M^3}{4} \sqrt{\frac{5}{\pi}} \frac{1}{(\rho^2+z^2)^{3/2}} \left[  \frac{3 z^2}{\rho^2+z^2} - 1 \right] \;.
\end{eqnarray}
We use Eq.\ (\ref{Weyl_rho}) and (\ref{Weyl_z}) to write $\psi_1^{l=2}$ in terms of Schwarzschild coordinates:
\begin{eqnarray}
	\psi_1^{l=2}(r,\theta) &=& \frac{B_2 M^3}{4} \sqrt{\frac{5}{\pi}} \frac{1}{d(r,\theta)^3} \left[\frac{3 (r-M)^2 \cos^2\theta}{d(r,\theta)^2} - 1\right] \;, \label{SchwBBH_l2_psi1}
\end{eqnarray}
where
\begin{eqnarray}
	d(r,\theta) \equiv (r^2-2Mr+M^2\cos^2\theta)^{1/2} \;. \label{drtheta}
\end{eqnarray}
The potential $\gamma_1$ that corresponds to this choice of $\psi_1$ is
\begin{eqnarray}
	\gamma_1(r,\theta) &=& B_2 \sqrt{\frac{5}{\pi}} \left[ \frac{(r-M)}{2} \frac{\left[ c_{20}(r) + c_{22}(r) \cos^2\theta \right]}{d(r,\theta)^5} - 1 \right] \;, \label{SchwBBH_l2_gamma1}
\end{eqnarray}
where
\begin{eqnarray}
	c_{20}(r) &=& 2(r-M)^4 - 5M^2(r-M)^2 + 3M^4 \;, \\
	c_{22}(r) &=& 5 M^2(r-M)^2-3M^4 \;.
\end{eqnarray}

We calculate the multipole moments for the bumpy Schwarzschild spacetime by following the procedure laid out in Sec.\ \ref{sec:GH_moments}. The first few multipole moments are listed below:
\begin{eqnarray}
	\delta \mathcal{M}_0 &=& 0 \;, \\
	\delta \mathcal{M}_1 &=& 0 \;, \\
	\delta \mathcal{M}_2 &=& -\frac{1}{2} B_2 M^3 \sqrt{\frac{5}{\pi}} \;, \\
	\delta \mathcal{M}_3 &=& 0 \;, \\
	\delta \mathcal{M}_4 &=& \frac{4}{7} B_2 M^5 \sqrt{\frac{5}{\pi}} \;, \\
	\delta \mathcal{M}_5 &=& 0 \;.
\end{eqnarray}
Thus an $l=2$ mass perturbation in the Weyl sector changes only the even Geroch-Hansen mass moments with $l\geq2$, and the perturbations to the moments depend on magnitude of the perturbation $B_2$.

\subsection{$l=3$ mass perturbation}
For an $l=3$ perturbation in the Weyl sector, the potentials $\psi_1$ and $\gamma_1$ are given by
\begin{eqnarray}
	\psi_1^{l=3}(r,\theta) &=& \frac{B_3 M^4}{4} \sqrt{\frac{7}{\pi}} \frac{1}{d(r,\theta)^4} \left[ \frac{5 (r-M)^3 \cos^3\theta}{d(r,\theta)^3} - \frac{3(r-M)\cos\theta}{d(r,\theta)} \right] \;, \label{SchwBBH_l3_psi1} \\
	\gamma_1^{l=3}(r,\theta) &=& \frac{B_3M^5}{2} \sqrt{\frac{7}{\pi}} \cos\theta \left[ \frac{ c_{30}(r) + c_{32}(r)\cos^2\theta + c_{34}(r)\cos^4\theta + c_{36}(r)\cos^6\theta }{d(r,\theta)^7} \right] \;, \label{SchwBBH_l3_gamma1}
\end{eqnarray}
where
\begin{eqnarray}
	c_{30}(r) &=& -3r(r-2M) \;,\\
	c_{32}(r) &=& 10r(r-2M) + 2M^2 \;,\\
	c_{34}(r) &=& -7r(r-2M) \;,\\
	c_{36}(r) &=& -2M^2 \;.
\end{eqnarray}

The first few multipole moments are listed below:
\begin{eqnarray}
	\delta \mathcal{M}_0 &=& 0 \;, \\
	\delta \mathcal{M}_1 &=& 0 \;, \\
	\delta \mathcal{M}_2 &=& 0 \;, \\
	\delta \mathcal{M}_3 &=& -\frac{1}{2} B_3 M^4 \sqrt{\frac{7}{\pi}} \;, \\
	\delta \mathcal{M}_4 &=& 0 \;, \\
	\delta \mathcal{M}_5 &=& \frac{2}{3} B_3 M^6 \sqrt{\frac{7}{\pi}} \;.
\end{eqnarray}
An $l=3$ mass perturbation in the Weyl sector changes only the odd Geroch-Hansen mass moments with $l\geq3$ by an amount proportional to $B_3$.

\section{Perturbations to the mass moments: Kerr background}
\label{sec:KerrBBH_mass}
Now we repeat this procedure for a bumpy Kerr black hole with perturbed mass moments. We generate this spacetime by applying the Newman-Janis algorithm {\cite{NJ}} to the bumpy Schwarzschild spacetime, yielding {\cite{VH}}
\begin{eqnarray}
	\nonumber ds^2 &=& -(1+2\psi_1) \left(1-\frac{2Mr}{\Sigma}\right)dt^2 - \gamma_1 \frac{4a^2Mr\sin^2\theta}{\Delta\Sigma}\:dt\:dr - (1+2\psi_1-\gamma_1)\frac{4aMr\sin^2\theta}{\Sigma}\:dt\:d\phi \\
		\nonumber && + (1+2\gamma_1-2\psi_1) \frac{\Sigma}{\Delta} dr^2 + 2\gamma_1 \left[ 1 + \frac{2Mr (r^2+a^2)}{\Delta\Sigma} \right] a \sin^2\theta\:dr\:d\phi + (1+2\gamma_1-2\psi_1)\Sigma\:d\theta^2 \\
					&& + \left[ (r^2+a^2)^2 - a^2 \Delta\sin^2\theta + (\gamma_1-\psi_1) \frac{8a^2M^2r^2\sin^2\theta}{\Sigma-2Mr} - 2\psi_1\frac{\Sigma^2\Delta}{\Sigma-2Mr} \right] \frac{\sin^2\theta}{\Sigma} \:d\phi^2 \;,
	\label{KerrBBH}
\end{eqnarray}
where, as before, $\Delta = r^2-2Mr+a^2$ and $\Sigma = r^2+a^2\cos^2\theta$. This spacetime limits to the Kerr metric when $\psi_1 \to 0$ and $\gamma_1 \to 0$; in the limit $a \to 0$, it reproduces the bumpy Schwarzschild metric.

As with the bumpy Schwarzschild case, we need to compute the conformal factor and the Ernst potential for the spacetime. The conformal factor is
\begin{eqnarray}
	\Omega &=& \bar{R}^2 \left[ \left(1-\frac{M^2-a^2}{4}\bar{R}^2\right)^2 - a^2\bar{R}^2\sin^2\theta \right]^{-1/2} (1-\gamma_1) \;.
\end{eqnarray}
To compute the Ernst potential, we need to compute the norm and the twist of the timelike Killing vector. The norm is straightforward; it is given by
\begin{eqnarray}
	\lambda &=& \left( 1 - \frac{2Mr}{\Sigma}  \right) (1+2\psi_1) \;. \label{SchwBBH_mass_lambda}
\end{eqnarray}
Notice that the portion of the norm proportional to the perturbation falls off as $r^{-(l+1)}$.

The twist is much trickier. Formally, it does not exist; the construction detailed in Ref.\ {\cite{VH}} for making a bumpy Kerr black hole does not leave the spacetime vacuum in GR. Consider, for example, the $l=2$ perturbation, described in more detail below. If we compute the Einstein tensor and enforce the Einstein equation $G_{\mu\nu} = 8\pi T_{\mu\nu}$, we find that, in the large $r$ limit and to leading order in $a$, the spacetime has a stress-energy tensor whose only non-vanishing components are
\begin{eqnarray}
	T_{\theta\phi} = \frac{3}{8\pi} a B_2 M^4 \sqrt{\frac{5}{\pi}} \frac{(3-5\cos^2\theta) \cos\theta \sin^3\theta}{r^5}  \;.
\end{eqnarray}
It may be possible to generalize Geroch-Hansen moments to nonvacuum spacetimes, but that is beyond the scope of this paper. For our purposes, it is sufficient to say that for the kinds of perturbations we have considered, the spacetime approaches vacuum very rapidly in the region where we need to use the twist. In general, since the stress-energy tensor is constructed from two derivatives of the metric, an order $l$ perturbation to the metric produces non-zero stress-energy tensor which fall off as $r^{-(l+3)}$. This enables us to define the spacetime's twist to the order necessary to define the Geroch-Hansen moments to order $l+1$.

We divide the curl into two parts: the gradient of a scalar function $\omega'$ plus some correction term $\upsilon_\alpha$:
\begin{eqnarray}
	\omega_\alpha &=& \nabla_\alpha \omega' + \upsilon_\alpha \;. \label{twist_correction}
\end{eqnarray}
For an order $l$ mass perturbation, the scalar function $\omega'$ is the same as for the unperturbed Kerr spacetime:
\begin{eqnarray}
	\omega' &=& -\frac{2aM\cos\theta}{\Sigma} \;.
\end{eqnarray}
The correction term $\upsilon_\alpha$ falls off as $r^{-(l+3)}$. Since, at large $r$, the portion of the norm proportional to the perturbation falls off as $r^{-(l+1)}$, the correction to the curl can be neglected. We thus treat $\omega'$ as the spacetime's twist. 

\subsection{$l=2$ mass perturbation}
We generate an $l=2$ mass moment perturbation on a Kerr background by applying the Newman-Janis algorithm to the potentials on a Schwarzschild background, Eq.\ (\ref{SchwBBH_l2_psi1}) and (\ref{SchwBBH_l2_gamma1}). This yields
\begin{eqnarray}
	\psi_1^{l=2}(r, \theta) &=& \frac{B_2M^3}{4}\sqrt{\frac{5}{\pi}} \frac{1}{d(r,\theta,a)^3}\left[\frac{3L(r,\theta,a)^2\cos^2\theta} {d(r,\theta,a)^2}- 1\right] \;, \\
\gamma_1^{l=2}(r, \theta) &=& B_2\sqrt{\frac{5}{\pi}} \left[ \frac{L(r,\theta,a)}{2} \frac{\left[c_{20}(r,a) + c_{22}(r,a)\cos^2\theta + c_{24}(r,a)\cos^4\theta\right]}{d(r,\theta,a)^5} - 1\right]\;,
\end{eqnarray}
where
\begin{eqnarray}
d(r, \theta, a) &=& \sqrt{r^2 - 2Mr + (M^2 + a^2)\cos^2\theta} \;, \label{drthetaa} \\
L(r,\theta,a) &=& \sqrt{(r - M)^2 + a^2\cos^2\theta}\;, \label{Lrthetaa}
\end{eqnarray}
and
\begin{eqnarray}
c_{20}(r,a) &=& 2(r-M)^4 - 5M^2(r-M)^2 + 3M^4 \;,\\
c_{22}(r,a) &=& 5M^2(r-M)^2 - 3M^4 + a^2\left[4(r-M)^2 - 5M^2\right] \;,\\
c_{24}(r,a) &=& a^2(2a^2 + 5M^2) \;.
\end{eqnarray}
The Ernst potential is given by
\begin{eqnarray}
	\xi = \frac{M}{r-M-ia\cos\theta} + \frac{B_2 M^3}{4} \sqrt{\frac{5}{\pi}} \frac{r(r-2M)}{(r-M)^2} \left[ \frac{r(r-2M) - (3r^2-6Mr+2M^2)\cos^2\theta}{d(r,\theta)^5} \right] \;.
\end{eqnarray}

We calculate the multipole moments following the procedure outlined in the previous sections. The changes to the multipole moments are
\begin{eqnarray}
	\delta \mathcal{M}_0 &=& 0 \;, \\
	\delta \mathcal{M}_1 &=& 0 \;, \\
	\delta \mathcal{M}_2 &=& -\frac{1}{2} B_2 M^3 \sqrt{\frac{5}{\pi}} \;, \\
	\delta \mathcal{M}_3 &=& 0 \;.
\end{eqnarray}
The lowest order multipole that is affected is the $l=2$ mass moment, and the change is the same as for an $l=2$ mass perturbation on a Schwarzschild background. The multipole moments are well-defined up to the $\mathcal{M}_3$ moment; the higher order moments are not well-defined because of the presence of a fluid that behaves in the large $r$ limit as an $l=4$ multipole.

\subsection{$l=3$ mass perturbation}
We generate an $l=3$ mass moment perturbation on a Kerr background by applying the Newman-Janis algorithm to the potentials on a Schwarzschild background, Eq.\ (\ref{SchwBBH_l3_psi1}) and (\ref{SchwBBH_l3_gamma1}), which yields
\begin{eqnarray}
	\psi_1^{l=3}(r, \theta) &=& \frac{B_3 M^4}{4} \sqrt{\frac{7}{\pi}} \frac{1}{d(r,\theta,a)^4} \left[ \frac{5 L(r,\theta,a)^3 \cos^3\theta}{d(r,\theta,a)^3} - \frac{3 L(r,\theta,a) \cos\theta}{d(r,\theta,a)} \right] \;, \\
\gamma_1^{l=3}(r, \theta) &=& \frac{B_3 M^5}{2} \sqrt{\frac{7}{\pi}} \cos\theta \left[ \frac{c_{30}(r,a) + c_{32}(r,a)\cos^2\theta + c_{34}(r,a)\cos^4\theta + c_{36}(r,a)\cos^6\theta}{d(r,\theta,a)^7} \right] \;,
\end{eqnarray}
where
\begin{eqnarray}
c_{30}(r,a) &=& -3r(r-2M) \;,\\
c_{32}(r,a) &=& 10r(r-2M) + 2M^5 - 3a^2 \;,\\
c_{34}(r,a) &=& -7r(r-2M) + 10a^2 \;, \\
c_{36}(r,a) &=& -2M^2 - 7a^2 \;.
\end{eqnarray}
The Ernst potential is given by
\begin{eqnarray}
	\xi = \frac{M}{r-M-ia\cos\theta} + \frac{B_3 M^4}{4} \sqrt{\frac{7}{\pi}} \frac{r(r-2M)}{r-M} \left[ \frac{3r(r-2M) - (5r^2-10Mr+2M^2) \cos^2\theta}{d(r,\theta)^7} \right]  \;.
\end{eqnarray}

The changes to the multipole moments are
\begin{eqnarray}
	\delta \mathcal{M}_0 &=& 0 \;, \\
	\delta \mathcal{M}_1 &=& 0 \;, \\
	\delta \mathcal{M}_2 &=& 0 \;, \\
	\delta \mathcal{M}_3 &=& -\frac{1}{2} B_3 M^4 \sqrt{\frac{7}{\pi}} \;, \\
	\delta \mathcal{M}_4 &=& 0 \;.
\end{eqnarray}
The perturbation changes the $l=3$ mass multipole, and the change is the same as for an $l=3$ mass perturbation on a Schwarzschild background. The lower order multipoles are unchanged.

\section{Perturbations to the spin moments: Schwarzschild background}
\label{sec:SchwBBH_spin}
When we defined mass perturbations in Sec.\ \ref{sec:SchwBBH_mass} and \ref{sec:KerrBBH_mass}, we based our definition on the multipolar expansion of the gravitational potential in Newtonian gravity. We cannot do the same thing for spin perturbations because in Newtonian gravity, mass currents are not a source of the gravitational field. Instead, we create perturbations to the spin moments by adding an imaginary perturbation to the Ernst potential as defined in Eq.\ (\ref{Ernst_Xi}):
\begin{eqnarray}
	\Xi = \Xi_\mathrm{Schw} + i \Xi_1
\end{eqnarray}
where $\Xi_\mathrm{Schw} = r/M-1$ is the Ernst potential for the Schwarzschild spacetime.

When we constructed mass perturbations, we defined a set of perturbations that are proportional to spherical harmonic functions in the large $r$ limit by imposing the condition that the metric satisfy the Einstein equations to first order. We want to define a similar set of perturbations to the spin moments. As shown by Ernst \cite{Ernst}, enforcing the vacuum Einstein equations to first order gives the following constraint equation for $\Xi_1$:
\begin{eqnarray}
	\nabla^2\left(\frac{\partial^2 \Xi_1}{\partial r^2}\right) = 0 \;. \label{DXi1_Laplace}
\end{eqnarray}
As discussed in {\cite{Ernst}}, one class of solutions to this equation includes the linearized Kerr spacetime;  we discuss this in more detail in the appendix. We construct bumpy black holes by considering another class of solutions to Eq.\ (\ref{DXi1_Laplace}). Since $\left( \partial^2 \Xi_1 / \partial r^2 \right)$ satisfies Laplace's equation, we can define an order $l$ spin perturbation as one for which $\left( \partial^2 \Xi_1 / \partial r^2 \right)$ is an order $l$ spherical harmonic. We can relate the perturbation to the Ernst potential to changes in the timelike Killing vector field by inverting Eq.\ (\ref{Ernst_Xi}):
\begin{eqnarray}
	\epsilon &=& \frac{\Xi-1}{\Xi+1} \\
		&=& 1 - \frac{2M}{r} + i \frac{2M^2}{r^2}\Xi_1 \;.
\end{eqnarray}
The perturbation leaves the norm of the timelike Killing vector unchanged, but it changes the twist of the timelike Killing vector, which was formerly zero, to
\begin{eqnarray}
	\omega &=& \frac{2M^2}{r^2} \Xi_1 \;.
\end{eqnarray}

The metric now has a nonzero $g_{t\phi}$ component, which we denote $\sigma_1$:
\begin{eqnarray}
	ds^2 &=& -\left(1-\frac{2M}{r}\right)dt^2 + 2\sigma_1 \: dt\:d\phi + \left(1-\frac{2M}{r}\right)^{-1} dr^2 + r^2 \: d\theta^2 + r^2 \sin^2\theta \: d\phi^2 \;, \label{SchwBBH_spin}
\end{eqnarray}
where $\sigma_1$ is related to $\Xi_1$ by
\begin{eqnarray}
	\pder{}{r}\left(\frac{r \sigma_1}{r-2M}\right) &=& \frac{2M^2\sin\theta}{(r-2M)^2} \pder{\Xi_1}{\theta} \;, \label{dsigma1dr} \\
	\pder{\sigma_1}{\theta} &=& \frac{2M^2\sin\theta}{r} \left( 2\Xi_1 - r\pder{\Xi_1}{r} \right) \;. \label{dsigma1dtheta}
\end{eqnarray}
Equations (\ref{dsigma1dr}) and (\ref{dsigma1dtheta}) overdetermine $\sigma_1$; we will use Eq.\ (\ref{dsigma1dr}) to calculate $\sigma_1$ with the boundary conditions
\begin{eqnarray}
	\lim_{r\to\infty} \sigma_1 &=& 0 \;, \\
	\lim_{r\to\infty} \pder{\sigma_1}{r} &=& 0 \;.
\end{eqnarray}
Once we have the perturbed metric, we can calculate the multipole moments using the same procedure as in Sec.\ \ref{sec:SchwBBH_mass}.

\subsection{$l=2$ spin perturbation}
We define an $l=2$ spin perturbation by specifying a solution to Eq.\ (\ref{DXi1_Laplace}) that has the form of an $l=2$ spherical harmonic:
\begin{eqnarray}
	\frac{\partial^2 \Xi_1}{\partial r^2} &=& \frac{S_2 M}{4} \sqrt{\frac{5}{\pi}} \frac{1}{(\rho^2+z^2)^{3/2}} \left[ \frac{3 z^2}{\rho^2+z^2} - 1 \right] \;.
	\label{D2xi}
\end{eqnarray}
We can use Eq.\ (\ref{Weyl_rho}) and (\ref{Weyl_z}) to rewrite Eq.\ (\ref{D2xi}) in terms of Schwarzschild coordinates:
\begin{eqnarray}
	\frac{\partial^2 \Xi_1}{\partial r^2} &=& \frac{S_2 M}{4} \sqrt{\frac{5}{\pi}} \frac{1}{d(r,\theta)^3} \left[ \frac{3 (r-M)^2 \cos^2\theta}{d(r,\theta)} - 1 \right] \;,
\end{eqnarray}
where $d(r,\theta)$ is defined in Eq.\ (\ref{drtheta}). Integrating, and imposing the conditions $\partial\sigma_1/\partial r \to 0$ and $\sigma_1\to 0$ as $r\to\infty$ yields
\begin{eqnarray}
	\Xi_1(r,\theta) &=& \frac{S_2}{4} \sqrt{\frac{5}{\pi}} \frac{\left[ d(r,\theta)^2 - (r-M) d(r,\theta) + M^2\cos^2\theta \right]}{M d(r,\theta)} \;.
\end{eqnarray}
From the perturbation to $\Xi$, we can calculate the perturbation to the metric $\sigma_1$:
\begin{eqnarray}
	\sigma_1 &=& \frac{S_2M}{2} \sqrt{\frac{5}{\pi}} \cos\theta \left[ \frac{2 d(r,\theta)}{r} - \frac{r-M}{d(r,\theta)} - \left(1-\frac{2M}{r}\right) \right] \;.
\end{eqnarray}

The first few multipole moments are listed below:
\begin{eqnarray}
	\delta \mathcal{M}_0 &=& 0 \;, \\
	\delta \mathcal{M}_1 &=& 0 \;, \\
	\delta \mathcal{M}_2 &=& i \frac{1}{4} S_2 M^3 \sqrt{\frac{5}{\pi}} \;, \\
	\delta \mathcal{M}_3 &=& 0 \;, \\
	\delta \mathcal{M}_4 &=& -i \frac{1}{28} S_2 M^5 \sqrt{\frac{5}{\pi}} \;, \\
	\delta \mathcal{M}_5 &=& 0 \;.
\end{eqnarray}
The $l=2$ spin perturbation changes the even spin moments for $l\geq2$, but leaves the odd spin moments and all of the mass moments unchanged.

\subsection{$l=3$ spin perturbation}
Consider an $l=3$ spin perturbation:
\begin{eqnarray}
	\frac{\partial^2 \Xi_1}{\partial r^2} = \frac{S_3 M^2}{4} \sqrt{\frac{7}{\pi}} \frac{1}{(\rho^2+z^2)^2} \left[ \frac{5 z^3}{(\rho^2+z^2)^{3/2}} - \frac{3 z}{(\rho^2+z^2)^{1/2}} \right] \;.
\end{eqnarray}
Then the perturbation to the Ernst potential $\Xi_1$ is given by
\begin{eqnarray}
	\Xi_1 = i \frac{S_3}{12} \sqrt{\frac{7}{\pi}} \cos\theta \left[ 3 - \frac{(r-M) (3d(r,\theta)^2-M^2\cos^2\theta)}{d(r,\theta)^3} \right] \;.
\end{eqnarray}
This corresponds to a perturbation to the metric given by
\begin{eqnarray}
	\sigma_1 = \frac{S_3 M}{6} \sqrt{\frac{7}{\pi}} \left[ \left(1-\frac{2M}{r}\right)\frac{r^2(2r-3M)-3Mr(r-2M)\cos^2\theta-2M^3\cos^4\theta}{d(r,\theta)^3} - \frac{2r-M-3M\cos^2\theta}{r} \right] \;.
\end{eqnarray}
The first few multipole moments are listed below:
\begin{eqnarray}
	\delta \mathcal{M}_0 &=& 0 \;, \\
	\delta \mathcal{M}_1 &=& 0 \;, \\
	\delta \mathcal{M}_2 &=& 0 \;, \\
	\delta \mathcal{M}_3 &=& i \frac{1}{12} S_3 M^4 \sqrt{\frac{7}{\pi}} \;, \\
	\delta \mathcal{M}_4 &=& 0 \;, \\
	\delta \mathcal{M}_5 &=& -i \frac{1}{36} S_3 M^6 \sqrt{\frac{7}{\pi}} \;.
\end{eqnarray}
The $l=3$ spin perturbation changes only the odd Geroch-Hansen moments for $l\geq3$.

\section{Perturbations to the spin moments: Kerr background}
\label{sec:KerrBBH_spin}
We generate a Kerr spacetime with perturbed spin moments by applying the Newman-Janis algorithm to the Schwarzschild spacetime with perturbed spin moments, whose metric is given in Eq.\ (\ref{SchwBBH_spin}). This yields
\begin{eqnarray}
	\nonumber ds^2 &=& -\left(1-\frac{2Mr}{\Sigma}-\frac{4aMr}{\Sigma^2}\sigma_1\right)dt^2 - \left[\frac{4aMr\sin^2\theta}{\Sigma} + 4\sigma_1 \left(\frac{(r^2+a^2)2Mr}{\Sigma^2}+\frac{\Delta}{\Sigma}-\frac{\Delta}{2\Sigma-4Mr}\right)\right]\:dt\:d\phi \\
		&& + \frac{\Sigma}{\Delta} dr^2 + \Sigma\:d\theta^2 + \left[ (r^2+a^2)^2 - a^2 \Delta\sin^2\theta + \frac{4aMr}{\Sigma-2Mr}\sigma_1 \left(\Delta + \frac{2Mr(r^2+a^2)}{\Sigma}\right) \right]\:\frac{\sin^2\theta}{\Sigma}\:d\phi^2 \;.
\end{eqnarray}
The norm of the timelike Killing vector is
\begin{eqnarray}
	\lambda = 1-\frac{2Mr}{\Sigma} + \frac{4aMr}{\Sigma^2} \sigma_1(r,\theta) \;. \label{KerrBBH_spin_lambda}
\end{eqnarray}
As in Sec.\ \ref{sec:KerrBBH_mass}, we cannot define the twist because the spacetime is not vacuum. For example, in the large $r$ limit and expanding in $a$, the stress-energy tensor of the Kerr spacetime with an $l=2$ spin perturbation, which we define in the following section, has nonzero terms
\begin{eqnarray}
	T_{\theta\theta} &=& -\frac{39}{8\pi} a S_2 M^4 \sqrt{\frac{5}{\pi}} \frac{\cos\theta \sin^2\theta}{r^5} \;, \\
	T_{\phi\phi} &=& -\frac{27}{8\pi} a S_2 M^4 \sqrt{\frac{5}{\pi}} \frac{\cos\theta \sin^4\theta}{r^5} \;.
\end{eqnarray}
In general, an order $l$ spin perturbation creates nonzero terms in the stress-energy tensor that fall off like $r^{-(l+3)}$. As in Sec.\ \ref{sec:KerrBBH_mass}, this allows us to define the Geroch-Hansen moments up to order $l+1$. Unlike in the case of a mass perturbation on a Kerr background, we will find that the twist depends on the perturbation, so we cannot go further in our calculation without choosing a particular perturbation.

\subsection{$l=2$ spin perturbation}
We define an $l=2$ spin perturbation with the perturbation potential
\begin{eqnarray}
	\sigma_1(r,\theta) &=& -\frac{S_2 M}{2} \sqrt{\frac{5}{\pi}} \cos\theta \left[ \frac{M}{L(r,\theta,a)+M} \frac{L(r,\theta,a)+M\sin^2\theta}{d(r,\theta,a)} + \frac{L(r,\theta,a)-M-d(r,\theta,a)}{L(r,\theta,a)+M} \right] \;,
\end{eqnarray}
where $d(r,\theta,a)$ and $L(r,\theta,a)$ are defined in Eq.\ (\ref{drthetaa}) and (\ref{Lrthetaa}), respectively. Now we need to construct the Ernst potential. The norm of the timelike Killing vector is given by Eq.\ (\ref{KerrBBH_spin_lambda}). As discussed in Sec.\ \ref{sec:KerrBBH_mass}, we can write the curl of a nonvacuum spacetime in the form of Eq.\ (\ref{twist_correction}). We define $\omega'$ by ignoring all terms of order $a S_2$ in the curl. This yields
\begin{eqnarray}
	\omega' = -\frac{2aM\cos\theta}{\Sigma} - \frac{S_2M}{2} \sqrt{\frac{5}{\pi}} \frac{1}{r} \left[ \frac{r^2-2Mr+2M^2\cos^2\theta}{r d(r,\theta)} - \left(1-\frac{M}{r}\right) \right] \;.
\end{eqnarray}
In this case, $\upsilon_\alpha$ falls off like $r^{-5}$, so we are justified in treating $\omega'$ as the twist. The Ernst potential is
\begin{eqnarray}
	\xi = \frac{M}{r-M-ia\cos\theta} + i \frac{S_2M}{4} \sqrt{\frac{5}{\pi}} \frac{r(r-2M) - (r-M) d(r,\theta) + 2M^2\cos^2\theta}{(r-M)^2 d(r,\theta)} \;.
\end{eqnarray}

Applying the procedure from the previous sections gives the multipole moments:
\begin{eqnarray}
	\delta \mathcal{M}_0 &=& 0 \;, \\
	\delta \mathcal{M}_1 &=& 0 \;, \\
	\delta \mathcal{M}_2 &=& i\frac{1}{4} S_2 M^3 \sqrt{\frac{5}{\pi}} \;, \\
	\delta \mathcal{M}_3 &=& 0 \;.
\end{eqnarray}
The perturbation changes the $l=2$ spin moment by the same amount as an $l=2$ spin perturbation on a Schwarzschild background, and it leaves lower order moments unchanged.

\subsection{$l=3$ spin perturbation}
We consider an $l=3$ spin perturbation:
\begin{eqnarray}
	\nonumber \sigma_1(r,\theta) &=& \frac{S_3 M}{6} \sqrt{\frac{7}{\pi}} \left[\frac{L(r,\theta,a)-M}{L(r,\theta,a)+M}\frac{s_{30}(r,a) + s_{32}(r,a)\cos^2\theta + s_{34}(r,a)\cos^4\theta}{d(r,\theta,a)^3} \right. \\
		&& \left. - \frac{2 L(r,\theta,a) - M(3\cos^2\theta-1)}{L(r,\theta,a)+M} \right] \;,
\end{eqnarray}
where
\begin{eqnarray}
	s_{30}(r,a) &=& 2L(r,\theta,a)^3 + 3r(r-2M) + 2M^3 \;, \\
	s_{32}(r,a) &=& -3M(r(r-2M)-a^2) \;, \\
	s_{34}(r,a) &=& -2M^3-3a^2M \;.
\end{eqnarray}

Now we need to construct the Ernst potential. The norm of the timelike Killing vector is given by Eq.\ (\ref{KerrBBH_spin_lambda}). As discussed in Sec.\ \ref{sec:KerrBBH_mass}, we can write the curl of a nonvacuum spacetime in the form of Eq.\ (\ref{twist_correction}). We define $\omega'$ by ignoring all terms of order $a S_3$ in the curl. This yields
\begin{eqnarray}
	\omega' = -\frac{2aM\cos\theta}{\Sigma} + \frac{S_3 M^2}{6}\sqrt{\frac{7}{\pi}} \frac{\cos\theta}{r^2} \left[ \frac{(r-M) (3r^2-6rM+2M^2\cos^2\theta)}{d(r,\theta)^3} - 3 \right] \;.
\end{eqnarray}
In this case, $\upsilon_\alpha$ falls off like $r^{-6}$, so we are justified in treating $\omega'$ as the twist. The Ernst potential is
\begin{eqnarray}
	\xi = \frac{M}{r-M-ia\cos\theta} + i \frac{S_3 M^2}{12}\sqrt{\frac{7}{\pi}} \cos\theta \left[ \frac{3r(r-2M)(d(r,\theta)-r+M) + M^2\cos^2\theta (3d(r,\theta)-2r+2M)}{(r-M)^2 d(r,\theta)^3} \right] \;.
\end{eqnarray}

Applying the procedure from the previous sections gives the multipole moments:
\begin{eqnarray}
	\delta \mathcal{M}_0 &=& 0 \;, \\
	\delta \mathcal{M}_1 &=& 0 \;, \\
	\delta \mathcal{M}_2 &=& 0 \;, \\
	\delta \mathcal{M}_3 &=& i \frac{1}{12} S_3 M^4 \sqrt{\frac{7}{\pi}} \;, \\
	\delta \mathcal{M}_4 &=& 0 \;.
\end{eqnarray}
As in the case of an $l=3$ spin perturbation on a Schwarzschild background, the perturbation changes the $l=3$ spin moment but leaves lower order moments unchanged.

\section{Conclusion}
\label{sec:conclusion}
This paper extends the formal definition of bumpy black holes that was introduced in {\cite{CH}} and {\cite{VH}}. We consider two kinds of static, axisymmetric perturbations: those that perturb the mass moments and those that perturb the spin moments. The mass perturbations, as defined in \cite{VH}, correspond to pure multipoles in the Weyl sector. We map these perturbations to changes in the Geroch-Hansen moments and showed that an order $l$ Weyl perturbation changes the Geroch-Hansen mass moments above $\mathcal{M}_l$ but leaves the lower order moments unchanged. We introduce bumps that perturb the spin moments, and we show that an order $l$ spin perturbation changes the Geroch-Hansen spin moments above $\mathcal{M}_l$. In principle, this allows us to build spacetimes whose multipoles agree with those of the Kerr spacetime up to some arbitrary order $L$ but differ for $l \geq L$.

Now that the bumpy black hole formalism has been developed, the next step is to use this framework to construct tests of general relativity. As discussed in {\cite{VH}}, changes to multipolar structure of the spacetime result in changes to the frequencies of geodesics. They also change the images that distant observers would see when rays are traced through the spacetime {\cite{PJ}}. Previous work has focused on the effect of perturbations to the mass moments; we expect that perturbations to the current moments will produce different effects. Detailed analysis will be needed to see if perturbed current moments also leave an observationally important imprint. For any system in which the relevant motions are well described by geodesics, at least on short timescales, this framework should enable us to construct useful and interesting tests of the black hole no-hair theorem.

\acknowledgments
The author thanks Scott Hughes and Leo Stein for useful discussions. The author also thanks Nico Yunes for helpful comments and suggestions. The program {\sc Mathematica} was used to perform the calculations. This work was supported by NSF Grant PHY-0449884 and by NASA grant NNX08AL42G.

\appendix
\section{Linearized Kerr}
One solution for a current-type perturbation is well known: the linearized Kerr spacetime, i.e. Kerr expanded to leading order in the spin parameter $a$. In our notation, this spacetime is given by Eq.\ (\ref{SchwBBH_spin}) with
\begin{eqnarray}
	\sigma_1 &=& -\frac{2aM\sin^2\theta}{r} \;.
\end{eqnarray}
The purpose of this appendix is to demonstrate that the framework we have developed includes the linearized Kerr spacetime; however, it requires a modification to the procedure we present in Sec.\ \ref{sec:SchwBBH_spin}.

To begin, note that one family of solutions to Eq.\ (\ref{DXi1_Laplace}) can be written
\begin{eqnarray}
	\Xi_1 = A f_0(\theta) + B r f_1(\theta) \;.
\end{eqnarray}
Let us choose a solution with $B=0$ and $f_0 = \cos\theta$. We calculate the perturbation to the spacetime $\sigma_1$ by enforcing Eq.\ (\ref{dsigma1dtheta}):
\begin{eqnarray}
	\frac{\partial\sigma_1}{\partial\theta} = \frac{4AM^2\sin\theta\cos\theta}{r}\;,
\end{eqnarray}
which we readily integrate to obtain
\begin{equation}
\sigma_1 = \frac{2AM^2\sin^2\theta}{r}\;.
\end{equation}
It is simple to verify that this satisfies Eq.\ (\ref{dsigma1dr}). If we now choose
\begin{equation}
A = -\frac{a}{M}\;,
\end{equation}
we identify this solution as the linearized Kerr spacetime.


\begin{thebibliography}{99}

\bibitem{Geroch} R.\ Geroch, J.\ Math.\ Phys. {\bf 11}, 2580 (1970).

\bibitem{Hansen} R.\ O.\ Hansen, J.\ Math.\ Phys. {\bf 15}, 46 (1974).

\bibitem{nohair} W.\ Israel, Phys.\ Rev.\ {\bf 164}, 1776 (1967); B.\ Carter, Phys.\ Rev.\ Lett.\ {\bf 26}, 331 (1971); D.\ C.\ Robinson, Phys.\ Rev.\ Lett.\ {\bf 34}, 905 (1975).

\bibitem{Ryan} F.\ D.\ Ryan, Phys.\ Rev.\ D {\bf 52}, 5707 (1995).

\bibitem{CH} N.\ A.\ Collins and S.\ A.\ Hughes, Phys.\ Rev.\ D {\bf  
  69}, 124022 (2004).

\bibitem{GB} K.\ Glampedakis and S.\ Babak, Class.\ Quantum Grav.\
  {\bf 23}, 4167 (2006).

\bibitem{VH} S.\ J.\ Vigeland and S.\ A.\ Hughes, Phys.\ Rev.\ D {\bf 81}, 024030 (2010).

\bibitem{NJ} E.\ T.\ Newman and A.\ I.\ Janis, J.\ Math.\ Phys. {\bf{6}}, 915 (1965).

\bibitem{Thorne1980} K.\ S.\ Thorne, Rev.\ Mod.\ Phys. {\bf 52}, 299 (1980).

\bibitem{YP} N.\ Yunes and F.\ Pretorius, Phys.\ Rev.\ D {\bf 79}, 084043 (2009).

\bibitem{SY} C.\ F.\ Sopuerta and N.\ Yunes, Phys.\ Rev.\ D {\bf 80}, 064006 (2009).

\bibitem{BH} T.\ B\"ackdahl and M.\ Herberthson, Class.\ Quant.\ Grav. {\bf 22}, 3585 (2005).

\bibitem{Ernst} F.\ J.\ Ernst, Phys.\ Rev. {\bf 167}, 1175 (1968).

\bibitem{Geroch1971} R.\ Geroch, J.\ Math.\ Phys. {\bf 12}, 918 (1971).

\bibitem{FHP} G.\ Fodor, C.\ Hoenselaers, Z.\ Perj\'es, J. Math. Phys. {\bf 30}, 10 (1989).

\bibitem{PJ} T.\ Johannsen and D.\ Psaltis, Astrophys.\ J. {\bf 716}, 187 (2010).

\end{thebibliography}
\end{document}